\documentclass[aps,pr,reprint,showpacs,superscriptaddress]{revtex4-2}
\usepackage{chngcntr}
\usepackage{float}
\usepackage{color}
\usepackage{tabularx}
\usepackage{graphicx}
\usepackage{dcolumn}
\usepackage{bm}
\usepackage{mathrsfs}
\usepackage[ruled,vlined]{algorithm2e}
\usepackage{array} %
\usepackage{amssymb}
\usepackage[utf8]{inputenc}
\usepackage{tabstackengine}
\setstackEOL{\cr}
\usepackage{amsfonts,amsmath,mathrsfs,booktabs}
\usepackage{multirow,times}
\usepackage{float,color,bm}
\usepackage{empheq}
\usepackage{hhline}
\usepackage{tikz}
\usepackage{lipsum,mathtools}
\usepackage{diagbox}
\usepackage{bbm}
\usepackage[colorlinks,
linkcolor=blue,
anchorcolor=blue,
urlcolor=blue,
citecolor=blue
]{hyperref}
\newlength{\halfpagewidth}
\divide\halfpagewidth by 1

\newcommand{\splitatcommas}[1]{%
	\begingroup
	\ifnum\mathcode`,="8000
	\else
	\begingroup\lccode`~=`, \lowercase{\endgroup
		\edef~{\mathchar\the\mathcode`, \penalty0 \noexpand\hspace{0pt plus 0.3em}}%
	}\mathcode`,="8000
	\fi
	#1%
	\endgroup
}

\newcommand{\RNum}[1]{\uppercase\expandafter{\romannumeral #1\relax}} 

\newcommand{\PreserveBackslash}[1]{\let\temp=\\#1\let\\=\temp}
\newcolumntype{C}[1]{>{\PreserveBackslash\centering}p{#1}}
\newcolumntype{R}[1]{>{\PreserveBackslash\raggedleft}p{#1}}
\newcolumntype{L}[1]{>{\PreserveBackslash\raggedright}p{#1}}
 \normalsize

\allowdisplaybreaks[3]
\begin{document}
	\title{Optimal coordination of resources: A solution from reinforcement learning}
	\author{Guozhong Zheng}
	\address{School of Physics and Information Technology, Shaanxi Normal University, Xi'an, 710062, P. R. China}
	\author{Weiran Cai}
	\address{School of Computer Science, Soochow University, Suzhou 215006, P. R. China}
	\author{Guanxiao Qi}
	\address{Institute of Neuroscience and Medicine, INM-10, Research Centre J\"ulich, J\"ulich, Germany.}
	\author{Jiqiang Zhang}\email{zhangjq13@lzu.edu.cn }
	\address{School of Physics, Ningxia University, Yinchuan, 750021, P. R. China}
	\author{Li Chen}\email{chenl@snnu.edu.cn}
	\address{School of Physics and Information Technology, Shaanxi Normal University, Xi'an, 710062, P. R. China}
	
	\date{\today}
\begin{abstract}
Efficient allocation is important in nature and human society, where individuals frequently compete for limited resources. The Minority Game (MG) is perhaps the simplest toy model to address this issue. However, most previous solutions assume that the strategies are provided \emph{a priori} and static, failing to capture their adaptive nature. Here, we introduce the reinforcement learning (RL) paradigm to MG,  where individuals adjust decisions based on accumulated experience and expected rewards dynamically. We find that this RL framework achieves optimal resource coordination when individuals balance the exploitation of experience with random exploration. Yet, the imbalanced strategies of the two lead to suboptimal partial coordination or even anti-coordination. Our mechanistic analysis reveals a symmetry-breaking in action preferences at the optimum, offering a fresh solution to the MG and new insights into the resource allocation problem.
\end{abstract}
\maketitle
\section{Introduction}\label{introduction}
Scarcity is the fundamental property in most economic problems~\cite{Samuelson2005Economics}, where our society is incapable of fulfilling the ever-increasing wants and needs in a world of finite resources, and is the root of most wars. This challenge of efficient resource allocation hence underpins most economic concerns. However, it differs from typical optimization problems; instead of a single objective function to maximize, resource allocation often requires Pareto-optimal solutions~\cite{Hwang2012multiple}, where improvements for one party inevitably diminish outcomes for others due to conflicting interests.

A pivotal insight from classical Economics is that markets, through self-organization, can reach an optimal allocation.
This is well captured by Adam Smith~\cite{Smith2003wealth} in his famous quote ``It is not from the benevolence of the butcher, the brewer, or the baker, that we expect our dinner, but from their regard to their own interest." The key question to be addressed here is, \emph{in the pursuit of self-interests for individuals, how the optimal allocation is achieved for the whole population?} While the general equilibrium theory~\cite{Arrow1974general} in Economics has provided foundational insights into the properties of optimal allocation, it fails to understand how such optimum emergence is achieved and under what conditions.

Significant advances came from the complexity science perspective~\cite{arthur1999complexity}, specifically, through the EI Farol bar problem~\cite{Arthur1994Inductive}, which was later formulated as the Minority Game (MG)~\cite{Challet1997Emergence,Challet2005MinorityGI}. The scenario of MG: an odd number $N$ of agents make repeated choices of two options, and those who are within the less crowded group win, while the others lose. The system is intrinsically frustrated because no solution can satisfy everyone. 
In their seminal work~\cite{Challet1997Emergence}, agents rely on pre-defined strategies drawn from a common pool, leading to various coordination phenomena, including herding or independent decisions, depending crucially on the strategy pool’s size and similarity among chosen strategies.

Numerous studies have since explored coordination mechanisms in MG.
Along one thread, statistical physicists utilize replica calculus combining with partition function ~\cite{Marsili2000exact,Marsili2001continuum,Hart2001acrowd} or generating functionals~\cite{heimel2001generating,de2011non,galla2003dynamics} to establish a connection to nonequilibrium phase transitions~\cite{chakraborti2015statistical,galla2008transition,bottazzi2002adaptive}. Several extensions also have been introduced and investigated by taking into account factors such as payoffs~\cite{li2000minority,liaw2007three}, strategy distributions~\cite{coolen2008generating,garrahan2001correlated}, and learning algorithms~\cite{montague2006imaging,catteeuw2011heterogeneous}, etc.
The other thread applies Boolean game dynamics, assuming possession of local information of the neighborhood~\cite{paczuski2000self,huang2012emergence,dyer2008consensus,zhang2013controlling,zhang2016controlling,zhou2005self}.
However, these models often rely on static rules and pre-designed strategies, which lack the adaptability observed in real-world decision-making.

Here, we turn to the paradigm of reinforcement learning (RL)~\cite{Sutton2018reinforcement} to provide a different solution to the MG. As one main category of machine learning algorithms, RL specializes in decision-making in complex scenarios and has shown its great potential in autonomous driving, natural language processing, gaming~\cite{Silver2016mastering}, etc. The design of RL was motivated by the behavior modes observed across different species and has a solid foundation in neuroscience~\cite{Lee2012neural, Rangel2008framework}. Recently, the paradigm of RL has been borrowed to study problems in evolutionary game theory, such as cooperation~\cite{Tanabe2012Evolution,Zhang2020understanding, Jia2021Local, Shi2022analysis, He2022migration,Song2022reinforcement, Wang2022Levy, Ding2023emergence,Wang2023Synergistic,Yang2024Interaction, Sheng2024catalytic,Zhao2024Emergence,Wang2024Mathematics}, trust~\cite{Zheng2024decoding}, fairness~\cite{zheng2024decodingfair}, resource allocation~\cite{Andrecut2001q, Zhang2019reinforcement}, collective motion~\cite{Incera2020Development,Wang2023Modeling}, and other collective behaviors of humans~\cite{Zhang2020oscillatory, Tomov2021multi, He2022q}. 
Till now, a few studies have tried to apply RL to study MG. Ref.~\cite{Andrecut2001Qlearning} presents the earliest attempt by adopting Q-learning to MG, where the herding effect is found to be suppressed. Similar observations with Q-learning are made in a larger population in Ref.~\cite{Si-Ping2019Reinforcement}, where they claimed that the population evolves toward the optimal allocation state but with persistent large fluctuations. These fluctuations are so large that it's not convincing that it's a satisfactory solution. The reason might come from the self-regarding setup where only the focal player's own action is considered, while all others' action information is neglected in decision-making. The key questions of whether the optimal allocation is learnable, and what new insights of RL can provide to the solution of the MG remain open.

In this work, we adopt a Q-learning algorithm~\cite{Watkins1989learning,Watkins1992Q} to solve MG, where each agent has a Q-table to guide one's decision-making. 
Our method differs from other works in that we adopt the other-regarding setup~\cite{zheng2024Evolution}, where different outcomes in the last round are specified to identify different states precisely.
Besides, a temperature-like parameter is introduced to control the tradeoff between the exploitation of the accumulated experience and the random exploration.
Surprisingly, within our proposed framework, a rich spectrum of collective behaviors is observed, including partial, optimal, and anti-coordination.
In the exploitation-only scenario, the population is prone to be trapped in the local minimums in the form of periodic states. The presence of an appropriate level of exploration acts as perturbation helping to escape the local minima, so that the optimal coordination as the global minimum is reached. However, too much exploration ruins the stability of optimal coordination and could lead to anti-coordination.
These findings are shown to be robust and the impact of the learning parameters is systematically discussed.

\begin{figure*}[htpb]
\centering
\includegraphics[width=0.8\linewidth]{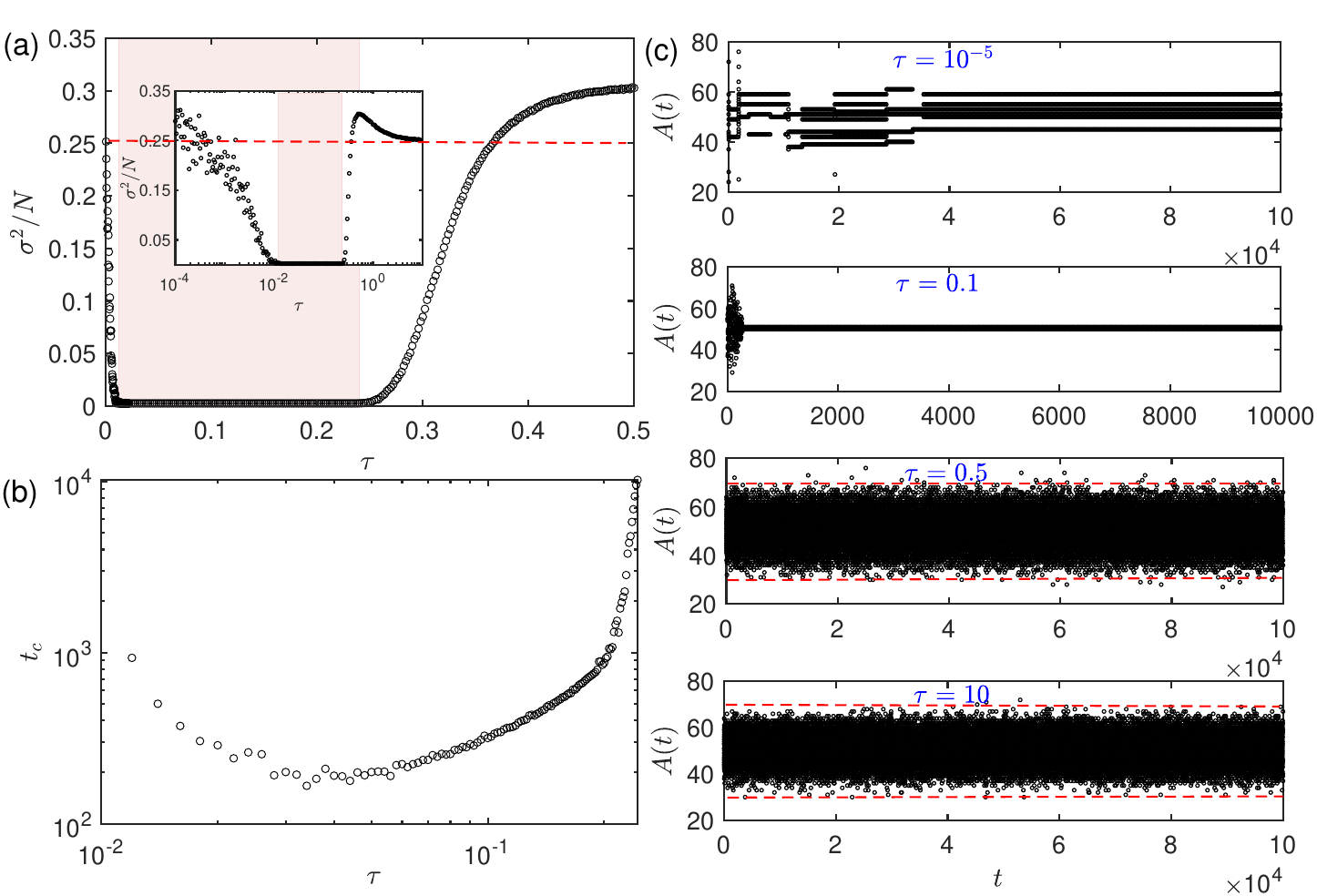}
\caption{\textbf{Emergence of coordination.} 
(a) The volatility as a function of the temperature $\tau$, and a higher temperature means the random exploration is more likely to be chosen than exploitation. Each data is averaged over 100 realizations, and for $5\times10^3$ time average after a transient of $5\times10^4$. The inset shows the plot with a logarithmic operation of $x$-axis, the shadowed zone corresponds to the region where the optimal coordination is achieved. The red dashed line ($\sigma^2/N=0.25$) corresponds to a benchmark scenario where individuals decide whether to go to the bar by simply flipping the coin.
(b) In the shadowed region ($\tau_{c_1}\le \tau \le\tau_{c_2}$, with $\tau_{c_1}\approx0.012$ and $\tau_{c_2}\approx0.24$), the average converging time $t_c$ towards the state of the optimal coordination versus the temperature $\tau$. Each data point is averaged over 300 realizations.
(c) Four typical time series of the number of people going to the bar $A(t)$ with the temperature $\tau=10^{-5}, 0.1, 0.5$, and 10 from the top to the bottom panels, which show the case of exploitation-only, OC, AC, and the exploration-only case, respectively. The two red dashed lines in the last two panels are the same ($A(t)=35, 70$) for comparison.
Parameters: $\alpha=0.1$, $\gamma=0.9$, $N=101$.
}
 \label{fig:tau}
 \end{figure*}

\section{Model}\label{sec:model}

Let's adopt the El Farol bar as our context.
The problem of the Minority Game is then rephrased as follows: Given an odd number of players, say $N$, in each round, every player has to independently make a binary choice -- go or not go to the bar. For simplicity, the bar capacity is assumed $C=N/2$. Thus, winners are those who end up on the minority side, and each gets 1 point. Losers are on the majority side and get zero.

Instead of distributing ready-made strategies to the individuals as in~\cite{Challet1997Emergence}, each has to learn a policy. 
Specifically, players adopt a model-free and value-based reinforcement learning -- the Q-learning algorithm~\cite{Watkins1989learning,Watkins1992Q}. Within this algorithm, each player has a Q-table in mind that directs the action choice of going or not going to the bar, see Table~\ref{tab:Qtable}. The states are the number of people who went to the bar in the last round denoted as $\mathbb{S}=\{s_1, ...,s_{N+1}\}$, and the action set includes two binary choices $\mathbb{A}=\{a_1, a_2\}$, corresponding to going or not going to the bar respectively. The elements $Q_{s,a}$ in Table~\ref{tab:Qtable} are the value function, measuring the value of action $a\in\mathbb{A}$ within the given state $s\in\mathbb{S}$. The action with a larger value of $Q_{s,a}$ is supposed to be more preferred within the given state $s$, according to the Q-learning algorithm.

The evolution follows a synchronous updating scheme, where every round includes two elementary processes -- the game and learning processes. Without loss of generality, all Q values in the table are set to zero, mimicking the status of no preference at the very beginning of the game. In round $t$, the first process is playing the Minority Game. To balance the trial-and-error exploration and the exploitation of the Q-table,  the softmax manner is employed by the player to choose an action, i.e., the probability of action $a_j$ being chosen for player $i$ is given by
\begin{equation}
p_{s^{i}, a_j}(t)=\frac{\exp \left(\frac{Q\left(s^{i}, a_j\right)}{\tau}\right)}{\sum\limits_{a_{k}\in\mathbb{A}} \exp \left(\frac{Q\left(s^{i}, a_k\right)}{\tau}\right)},
\label{eq:probability}
\end{equation}
where $s^{i}$ is the state of player $i$, and there are two actions in our case. Note that in the Minority Game, the states are identical for all players, i.e. $s^{i}\!=\!s(t)\!=\!A(t\!-\!1), \forall i\!\in\!\{1,2, ..., N\}$, $A(t\!-\!1)$ is the number of people who went to the bar at the end of round $t-1$.
$\tau$ is a temperature-like parameter that controls the trade-off between the exploration and exploitation events. While the exploitation is to follow the guidance of the Q-table, the exploration goes beyond the Q-table by trying random actions to explore the environment.
At a low temperature $\tau$, the action with a larger Q-value is probably to be chosen, whereas a random action choice is more likely to be selected for a larger $\tau$. In the extreme case of  $\tau\rightarrow0$, players strictly choose the action with the larger value, i.e. only exploitation of the Q-table. In the other extreme of $\tau\rightarrow+\infty$, players essentially make a random choice of the two actions irrespective of their Q-values, i.e. only the exploration is conducted. The ideal policies, however generally require a balance of the two, the optimal temperature $\tau$ is supposed to be located in between the two extreme values.
Once every player makes the choice of going or not going to the bar, the winning side can then be determined by comparing $A(t)$ with the capacity $C$, and also the reward $R$ for every player is obtained.

\begin{table}[tb]
\centering
\begin{tabular}{c|c|c}
\hline
\diagbox{State}{Action} & Go ($a_{1}$) & Not go ($a_{2}$) \\ \hline
0 $(s_{1})$             & $Q_{s_{1},a_{1}}$ & $Q_{s_{1},a_{2}}$ \\
1 $(s_{2})$             & $Q_{s_{2},a_{1}}$ & $Q_{s_{2},a_{2}}$ \\
2 $(s_{3})$             & $Q_{s_{3},a_{1}}$ & $Q_{s_{3},a_{2}}$ \\
3 $(s_{4})$             & $Q_{s_{4},a_{1}}$ & $Q_{s_{4},a_{2}}$ \\
...                     & ...               & ...               \\
N $(s_{N+1})$           & $Q_{s_{N+1},a_{1}}$ & $Q_{s_{N+1},a_{2}}$ \\ \hline
\end{tabular}
\caption{Q-table for each individual, where the state corresponds to how many people went to the bar in the last round, which is identical for all players, and the actions are the two choices of going or not going to the bar.}
\label{tab:Qtable}
\end{table}

To this end, evolution enters the second process, where all players need to revise their Q-tables as learning. Specifically, for player $i$, the element in its Q-table is updated according to the Bellman equation as follows
\begin{equation}
\begin{split}
Q_{s^{i}, a^i}(t+1)&=\alpha\left(\gamma \sum\limits_{a_{j}\in\mathbb{A}}p_{s^{\prime}, a_j}(t)Q_{s^{\prime}, a_j}(t) + R\right)\\
&+(1-\alpha)Q_{s^{i}, a^i}(t), \label{eq:Bellman_eq}
\end{split}
\end{equation}
where $s^{i}=s(t)$ and $a^i$ are respectively the state and the action taken at round $t$, $s'=A(t)$ is the state for the next round $t+1$, $R$ is the reward of player $i$ in this round, equal 1 or 0.
$\alpha\in(0,1]$ is the learning rate that determines the evolution pace of the Q-table, a small $\alpha$ corresponds to a slow evolution, the old value is largely kept, and the historical experience is well preserved. A large value of $\alpha$ can be interpreted as being forgetful instead.
 $\gamma\in[0,1)$ is the discount factor, determines the importance of future rewards, as the associated term $\sum\limits_{a_{j}\in\mathbb{A}} p_{s^{\prime}, a_j}(t) Q_{s^{\prime}, a_j}(t)\equiv \bar{Q}(t+1)$ gives the expected value in the new round $t+1$. A larger value of $\gamma$ means that players pay more attention to the future's guidance, and have a long-term vision. After every player completes their learning processes, the evolution of round $t$ is finished. The two processes repeat until the population reaches equilibrium or the evolution time reaches the desired long durations. For clarity, the evolution procedure is summarized in the protocol flowchart in Fig.~\ref{fig:protocol} in Appendix~\ref{sec:one}.

 Unless otherwise stated, we fix $N=101$ and the bar capacity $C=N/2$. To measure the degree of utilization of the bar, we define the volatility~\cite{Challet2005MinorityGI} as $\sigma^2/N$, where $\sigma^2=\langle (A(t)-C)^2\rangle_t$, $A(t)$ is the number of people who went to the bar in round $t$, and $\langle \cdots \rangle_t$ represents the time average. The smaller the volatility the better utilized the bar. If $\sigma^2/N\rightarrow0$, the bar maximizes its utilization as $A(t)\approx C$.
As a benchmark, when individuals randomly choose to go or not to go to the bar, e.g. by flipping a coin, $\sigma^2/N\rightarrow0.25$ is expected~\cite{Challet2005MinorityGI}.

A key question in the RL paradigm is how to properly handle the exploration-exploitation dilemma~\cite{Sutton2018reinforcement}. On one hand, the scenario of exploitation-only strategy may fall into suboptimal solutions without sufficiently exploring the untried possibilities; on the other hand, the individuals within the exploration-only scenario neglect the lessons drawn from experiences, they decide simply by flipping the coin and certainly also fail coordination. Therefore, how to balance the two is the primary question to be addressed here.

\begin{figure*}[htpb]
 \centering
\includegraphics[width=0.8\linewidth]{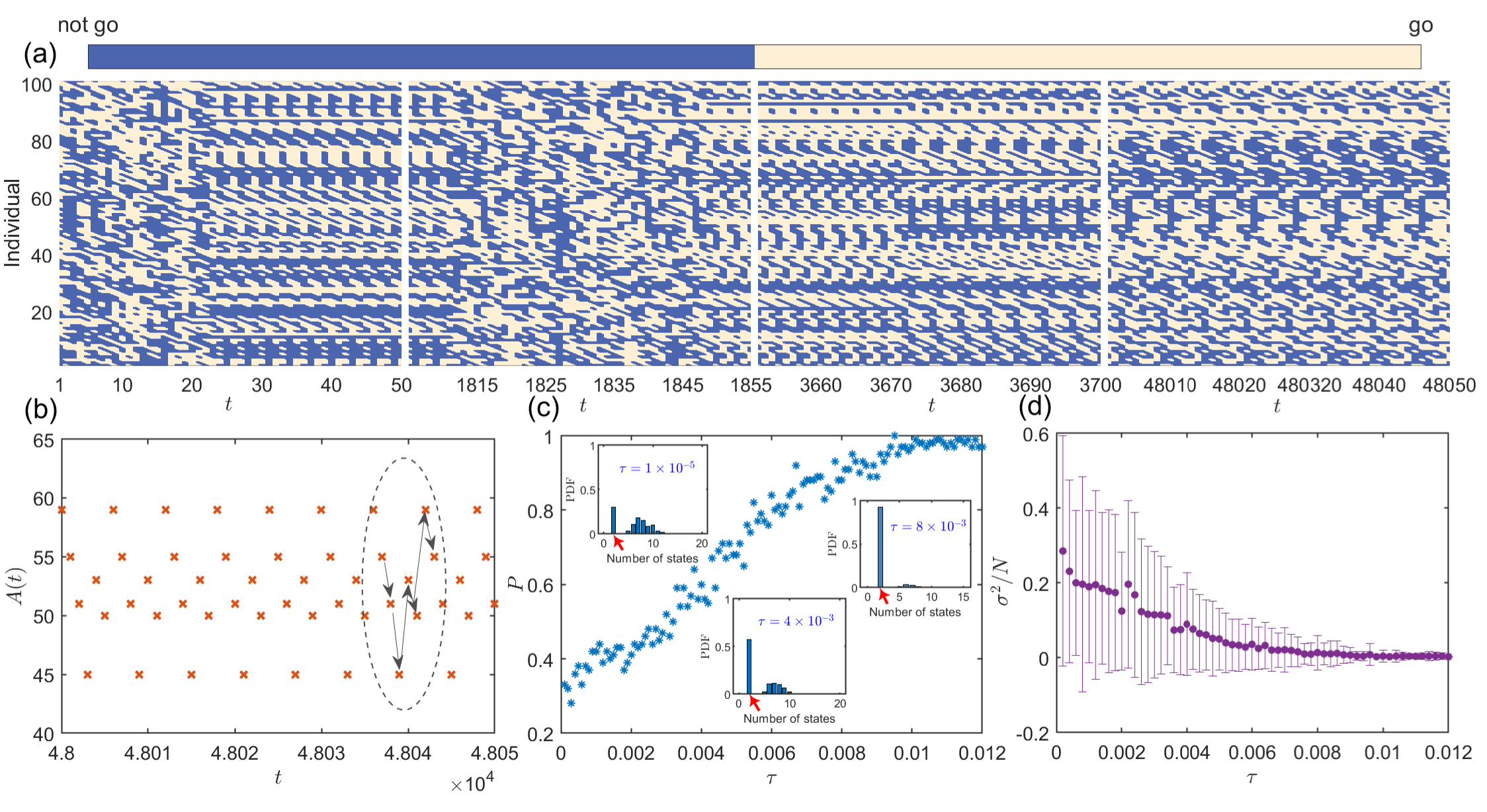}
\caption{\textbf{The evolution of the exploitation-only case.}
(a) The spatial-temporal patterns of action with the temperature $\tau=10^{-5}$, consisting of four typical time windows (from left to right): $0\sim50$, $1805\sim1855$, $3650\sim3700$, and $48000\sim48050$, they respectively show the formation of the periodic state for the first time, its collapse and reestablishment of new periodic states (the middle two panels), and a stable periodic state in the end.
(b) Time series of the attending number $A(t)$ at the end of (a), with a periodic evolution of $A(t)$ is encircled.
(c) The probability of falling into the optimal 50-51 state as a function of the temperature $\tau\in(0,0.012]$. The inset shows the histogram of the probability density distribution of different periods of stable states for three typical temperatures $\tau=1.0\times 10^{-5}$, $4.0\times 10^{-3}$ and $8.0\times 10^{-3}$, and the red arrow indicates the bar for the optimal state. 500 realizations are conducted for each $\tau$, and the state is sampled just after the transient of $1.0\times 10^{5}$ steps.
(d) The corresponding volatility as in (c), with the error bars representing the standard deviation.
Parameters: $\alpha=0.1$, $\gamma=0.9$, $N=101$.}
 \label{fig:exploitation_only}
 \end{figure*}

\section{Results}\label{sec:results}

We first report the impact of the temperature $\tau$ on the game evolution to examine the exploration-exploitation dilemma, where qualitatively distinct phases are observed, as shown in Fig.~\ref{fig:tau}.
Fig.~\ref{fig:tau}(a) shows the results of the volatility as a function of the temperature $\tau$ to measure the degree of bar utilization, which reveals a non-monotonic dependence as expected. As individuals are inclined to exploitation-only ($\tau\rightarrow0$), the volatility is as large as around 0.25, meaning that the coordination is failed and as bad as the benchmark scenario where players make decisions simply by flipping the coin. As $\tau$ increases ($\tau<\tau_{c_1}\approx0.012$), where the trial-and-error explorations are engaged, the volatility is reduced. This means that the fluctuations around the bar capacity $C$ are gradually reduced, and the population starts to learn how to coordinate to improve the utilization of the bar. The outcome is, however, not optimal, and thus can be termed as the~\emph{partial coordination} (PC) state.
Surprisingly, as $\tau$ continues to increase ($\tau_{c_1}\le\tau\le\tau_{c_2}\approx0.24$), the population enters into the~\emph{optimal coordination} (OC) state, where the attending number $A(t)$ fluctuates around $C$ in the 50-51 manner --- the optimal scenario one can expect, the volatility has reached its minimum.
As $\tau$ is further increased ($\tau>\tau_{c_2}$), however, the volatility starts to rise, and there is a region where $\sigma^2/N>0.25$, meaning the coordination is even worse than the benchmark scenario. We term this phenomenon as~\emph{anti-coordination} (AC). Finally, as individuals tend to be exploration-only ($\tau \gtrsim 5.0$), the volatility is approximately 0.25, equal to the benchmark value as expected.

Typical time series are shown in Fig.~\ref{fig:tau}(c), which respectively show the case of exploitation-only, OC, AC, and the exploration-only case from the top to the bottom. In the case of exploitation-only ($\tau=10^{-5}$), the number of people going to the bar $A(t)$ evolves into some meta-stable periodic states, but this state is not satisfactory as the value of $A(t)$ deviates considerably from the bar capacity $C$.
In the case of OC ($\tau=0.1$), the number of people going to the bar strictly fluctuates between the two cases of 50 and 51, the best solution one can expect in the MG.
As the temperature becomes large ($\tau=0.5, 10$), the evolution of $A(t)$ becomes random and is thus unsatisfactory as expected, but careful examination reveals that the fluctuations in the AC region ($\tau=0.5$) are even stronger than the extreme case ($\tau=10$).
Even though the solutions in the OC region ($\tau\in[\tau_{c_1},\tau_{c_2}]$) are all the same in the end, there the transient time $t_c$ differs for different temperatures $\tau$. Fig.~\ref{fig:tau}(b) shows that there exists the shortest converging time $t_c$ towards the optimal solution at around $\tau\approx0.034$. With this temperature, the population can reach the OC state within the shortest time.
Overall, the observations in Fig.~\ref{fig:tau} show that the population fails to coordinate in either extreme of exploitation-only or exploration-only,
 but they benefit from the trade-off between exploration and exploitation, and the optimal scenario emerges in this exploration-exploitation dilemma.

\section{Dynamical mechanism}\label{sec:analysis}

To understand how individuals succeed in coordination in some cases and why they fail in some other cases, we turn to the dynamical mechanism analysis. To this aim, here we fix the two learning parameters $\alpha=0.1$ and $\gamma=0.9$, where individuals appreciate both their historical experience and returns in the future.
\subsection{Exploitation-only}\label{subsec:first_transition_point}

We first focus on the exploitation-only case ($\tau\rightarrow0$), where the decision-making is strictly guided by one's Q-table.
A typical series of spatial-temporal patterns is shown in Fig.~\ref{fig:exploitation_only}(a). It shows that, after the transient, the population enters into some periodic states (the leftmost panel), but these periodic states are metastable, they become destabilized and the system evolves into other metastable states (the two panels in the middle). This process repeats until a stable periodic state is reached (the rightmost panel).

The dynamical mechanism behind this evolution is caused by the ``stubbornness" of individuals due to the small value of $\tau$, which makes the population prone to fall into metastable states.
To be specific, let's consider individual $i$ at round $t$, after it chooses action $a^i\in\mathbb{A}$ , it falls into two situations:
(i) If $i$ happens to be on the winning side, the reward $R=1$, this immediately makes the Q-value of the corresponding action $a^i$ larger than the opposite action denoted as $\tilde{a}^i$, i.e. $Q_{s^{i},a^i}>Q_{s^{i},\tilde{a}^i}$. This difference makes the individual firmly choose the action $a^i$ when the system enters the same state $s(t)$ next time, as the small value of $\tau$ can turn a small advantage in the Q-value into a strong preference in the action selection, as the two action probabilities $p_{s^{i},a^i}\rightarrow 1$ and $p_{s^{i},\tilde{a}^i}\rightarrow 0$ according to Eq.~(\ref{eq:probability}).
(ii) Otherwise, individual $i$ is on the losing side with $R=0$. In this situation, there are two different subsequent scenarios: 
(1) the new state $s(t+1)$ has not been experienced before, or the new state has been experienced but with two $Q$ values still being zero; in both cases, there the expected value $\bar{Q}(t+1)=0$. Therefore $Q_{s^{i},a^i}(t+1)=(1-\alpha)Q_{s^{i},a^i}(t)$, no preference is formed within the state $s(t)$.
(2) But if the new state was experienced and individual $i$ got positive rewards, $\bar{Q}(t+1)>0$, this contributes to the increase in $Q_{s^{i},a^i}$, the action $a^i$ is thus preferred next time when the system is within the state $s(t)$.
In brief, the ``stubbornness" of players is due to the extreme of $\tau\rightarrow 0$, whereby a small difference between $Q_{s^{i}, a_1}$ and $Q_{s^{i},a_2}$ is amplified and yields a strong preference in action selection. Therefore, one stubbornly chooses the action with the larger Q-value until this difference vanishes. Since the population is of finite size, a cycle of the $s(t)$ is easily formed by evolution, i.e., the population falls into some periodic states.

Within such a periodic state, each individual has a clear preference for each state at the early stage after the cycle is formed, the evolution becomes almost deterministic. But why do some periodic states become unstable? The reason lies in the fact that for individual $i$, even though the preferences are formed for each state $s^{i}$ within this circle; if no positive reward $R$ is obtained through the whole circle, the preference is weakened gradually for the decaying relationship $Q_{s^{i},a^i}(t+1)=(1-\alpha)Q_{s^{i},a^i}(t)$. Finally, when the two Q values become so close that there is no vanishing action probability for $p_{s^{i}, a_{1,2}}$, the stability of the cycle is broken as the choice of action becomes not deterministic anymore. Thereafter, the periodic state collapses, and pattern reconfiguration in the form of turbulence-like or quick arrangement may arise as shown in the second and third snapshots in Fig.~\ref{fig:exploitation_only}(a).
A periodic state is stable only when all individuals have at least one positive reward $R=1$ through a whole circle, whereby the expected value $\bar{Q}(t)>0$ in each state within the cycle, and all the preferences are strengthened to confront their decay. Such a stable state is seen in the last snapshot in Fig.~\ref{fig:exploitation_only}(a). The corresponding time series of the attending number $A(t)$ is shown in  Fig.~\ref{fig:exploitation_only}(b), which is a periodic state with period 6.

 \begin{figure}[tpb]
 \centering
\includegraphics[width=0.8\linewidth]{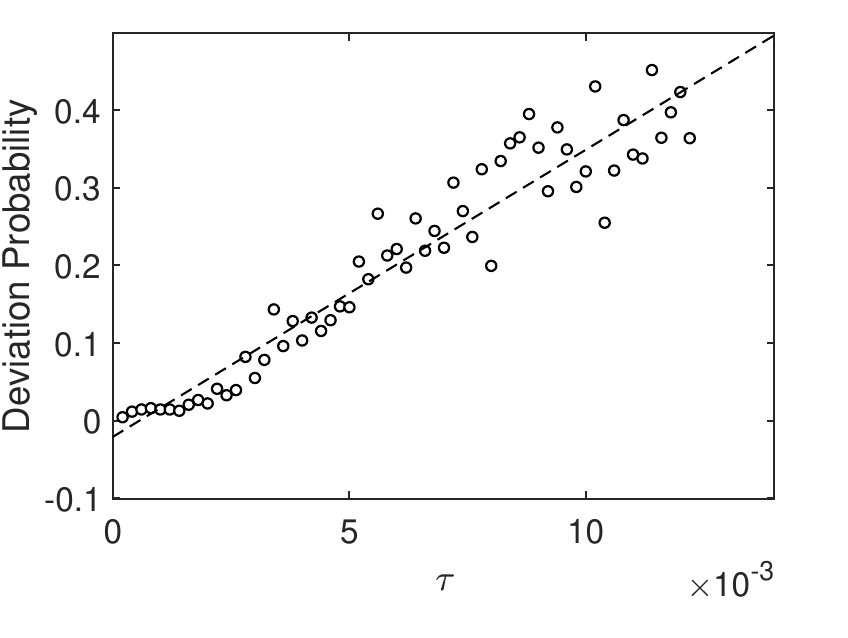}
\caption{\textbf{The deviation probability versus the temperature $\tau$}. 
The deviation probability is counted when the individuals select the action with the smaller Q-value, deviating from the guidance of the Q-table. The dashed line is a linear approximation. 100 ensemble averages are conducted for each $\tau$. Parameters: $\alpha=0.1$, $\gamma=0.9$, $N=101$.}
 \label{fig:deviation_prob}
 \end{figure}

\subsection{Partial coordination}\label{subsubsec:simulation_for_analysis}

As the temperature $\tau$ increases, the exploration events are engaged, and the evolution to the periodic states becomes harder and harder.
By contrast, the increase in $\tau$ prompts the probability of evolving toward the OC state shown in Fig.~\ref{fig:exploitation_only}(c), which provides the probability that the system finally evolves into the OC state as a function of $\tau$ with $\tau\in(0,0.012]$. We can see that even in the extreme case of $\tau\rightarrow 0$, there is already a probability around $30\%$ that the system falls into the OC state; as $\tau$ is increased, this probability continually increases and approaches 1 as $\tau\rightarrow 0.012$.
Detailed statistical analysis confirms this observation in insets of Fig.~\ref{fig:exploitation_only}(c), by showing the probability density function (PDF) of the number of states after the transient. We observe that some relatively long periodic states (such as periodic--7, 8) abound for small $\tau$ (e.g. $\tau=1.0\times 10^{-5}$), but they become fewer (e.g. $\tau=4.0\times 10^{-3}$), and almost disappear as $\tau$ becomes large (e.g. $\tau=0.008$), where the OC state dominates. But notice that the OC is not a periodic state, $A(t)$ randomly switches between $50$ and $51$.
Consequently, the resulting volatility also declines with the increase of $\tau$ in this range, approaching the minimum of 0.0025 for the OC state, shown in Fig.~\ref{fig:exploitation_only}(d). 
The shortened error bars indicate the decrease in the number of metastable states.

\begin{figure}
 \centering
 \includegraphics[width=0.8\linewidth]{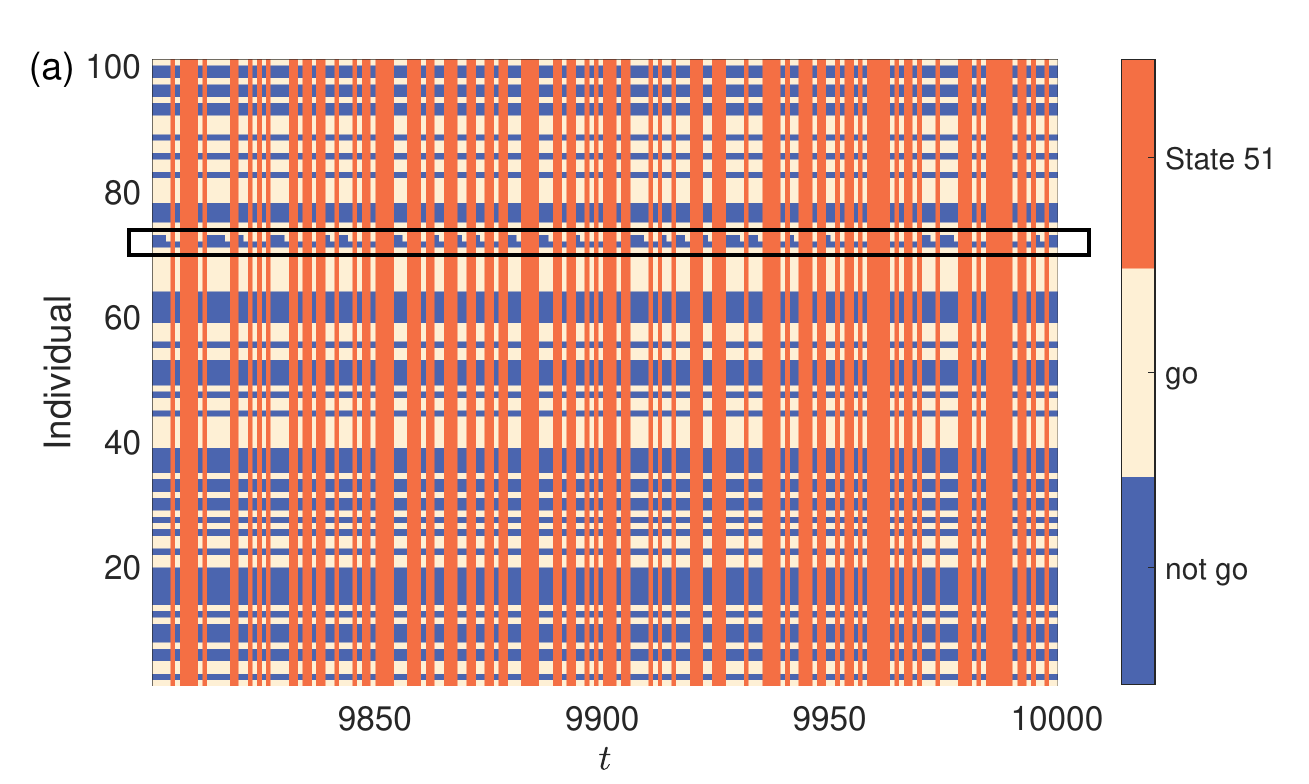}\\
 \includegraphics[width=0.8\linewidth]{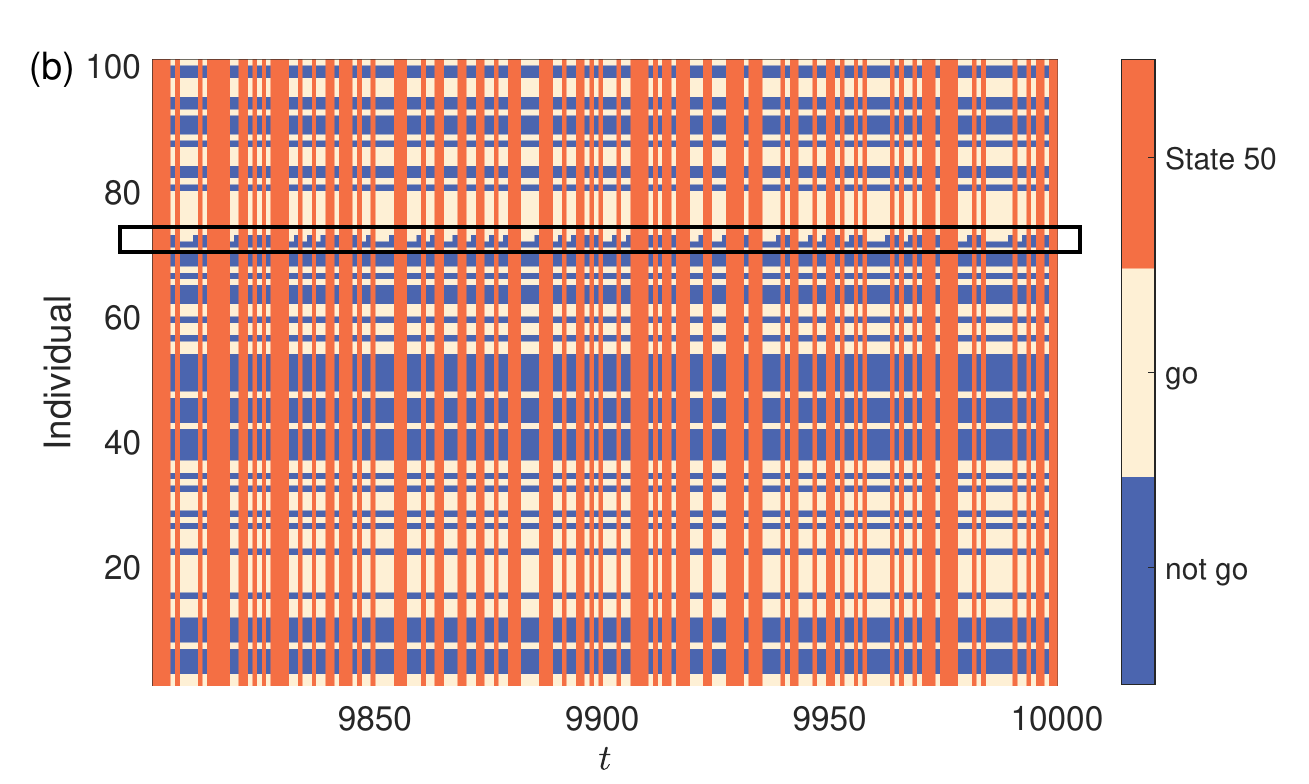}
 \caption{\textbf{The organization of optimal coordination}.
 The stationary patterns within the optimal coordination (OC) state, respectively in $s(t)=50$ (a) and $s(t)=51$ (b).
Besides the two actions (``go" and ``not go"), being within the other state is color-coded with orange.
In both states, an individual (here the one labeled ``71" indicated by a rectangle) emerges as the irresolute one who randomly switches its action, whereas all other individuals' actions remain unchanged all the time.
 Parameters: $\tau=0.1$, $\alpha=0.1$, $\gamma=0.9$, $N=101$.
 }
 \label{fig:OC}
 \end{figure}

According to Eq.~(\ref{eq:probability}), increasing $\tau$ increases the likelihood of choosing the action with the smaller Q-value, which can be interpreted as that individuals become less stubborn, as they deviate from the guidance of their Q-tables more frequently. As $\tau$ increases, this deviation probability increases, as shown in Fig.~\ref{fig:deviation_prob}.
From the physics point of view, those periodic states are suboptimal solutions and can be taken as local minimums in the phase space, and the temperature acts as the perturbation. With a small value of $\tau$, the population state is easily trapped in local minimums. Increasing $\tau$, the perturbations destabilize their stay in those local minimums, and the system escapes from local minimums and becomes more likely to fall into the global minimum --- the OC state.
As the temperature $\tau>\tau_{c_1}$, none of these periodic states is stable and the system falls into the OC state in all realizations.
An evolving pattern is shown in Appendix~\ref{sec:two}, where one can intuitively see the formation of the OC state and the temporal evolution of the associated fluctuations.

 \begin{figure*}
 \centering
\includegraphics[width=0.85\linewidth]{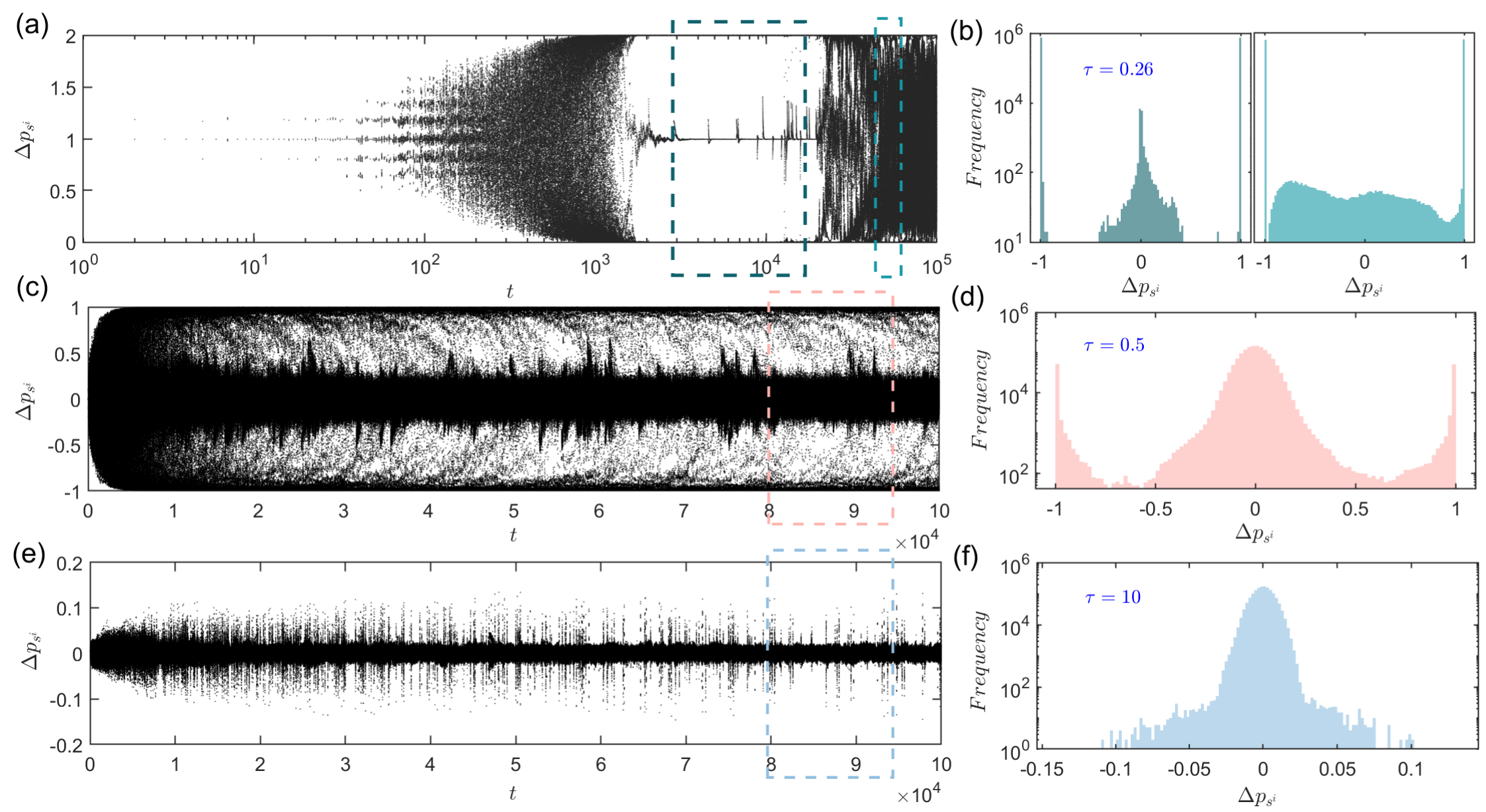}
\caption{\textbf{The evolution of preference behind the destabilization of OC state.}
The temporal evolution of preference defined as $\Delta p_{{s^{i}}}=p_{{s^{i}},{a_1}}-p_{{s^{i}},{a_2}}$ and its frequency within high-temperature scenarios: (a, b) $\tau=0.26$, (c, d) $\tau=0.5$, (e, f) $\tau=10$.
The frequencies are sampled in the rectangle at the corresponding left panel. The two panels in (b) are, respectively, for the period of optimal coordination and after its destabilization.
Notice that, the $x$-axis in (a)  and $y$-axis in (b, d, f) are in logarithmic form.
Other parameters: $\alpha=0.1$, $\gamma=0.9$, $N=101$.}
 \label{fig:delta_p}
 \end{figure*}

\subsection{Optimal coordination}

But how is the OC state orchestrated at the individual level?  Our analysis shows that within either state $A(t)=50$ or $51$, there is a spontaneous symmetry-breaking in two action choices, the population is self-organized into two subgroups: 50 individuals are determined to go to the bar, 50 choose not to go, leaving one irresolute individual who randomly switches between the two choices. This can be seen from the action patterns within the OC state shown in Fig.~\ref{fig:OC}, where the state has been stationary.

Fig.~\ref{fig:OC}(a) shows the case within the state $s(t)=50$. We can see that all individuals except the one labeled ``71" in this example keep consistent action choices. Specifically, $50$ individuals stick to going to the bar, and the other $50$ ones stick to not going to the bar. This leaves an awkward situation for individual 71: whichever side it chooses, the chosen side becomes 51 and is thus the losing side. The pathetic fact for individual 71 is that the same evolution repeats within the state $s(t)=50$ and it always gets zero rewards. The associated two Q-values $Q_{50,a_{1,2}}\approx 0$, the two actions are selected with an equal probability, which explains the random choice in the OC state. For the rest $N-1$ individuals, the expected reward $\langle R\rangle=1/2$ within $s(t)=50$ for each, which in turn strengthens their preferences with $s(t)=50$ in their Q-tables (see Appendix~\ref{sec:three}).
Similar observations are made within the state $s(t)=51$, where individual 71 still acts as the pathetic one, see Fig.~\ref{fig:OC}(b).
Note that the emergence of this pathetic individual in either state is purely by chance, whose preference is the least determined at the early stage of evolution. This makes the pathetic individual who restlessly switches her/his actions, and the continuous failure ($R=0$) leads to one's indecisiveness. By contrast, the continuing return for the rest ($\langle R\rangle=1/2$) strengthens their preferences in the two states. Therefore, it is the pathetic individual who determines which of the two subgroups wins, and its action switch guarantees the frozen preferences of the rest, which is crucial for the stability of the OC state. It is worth noting that before reaching this stationary state, the pathetic individuals are generally different within the two states, their competition turns the less rewarded one into the pathetic individual in both states. This nontrivial dynamical process is discussed in  Appendix~\ref{sec:four}.

What if the determined players made a mistake? Does such an event ruin the organization of the OC state? The answer is no. By carefully examining the impact of such events, we find that they indeed occur in the OC state. This may make the pathetic individual obtain a reward of 1 with a probability of $50\%$, and consequently, an action preference may be formed as its two Q-values are not equal anymore. However, the action preference of the pathetic individual is fairly weak compared to the determined players, including the one who made the mistake. In later evolution, the pathetic individual is still on the losing side, and the action preference is fading away as time goes by. Therefore, the role of all players remains unchanged, and thus the stability of the OC state is guaranteed.

\subsection{Anti-coordination}\label{subsec:AC}

As the temperature $\tau$ further increases, the OC state is ruined and the anti-coordination may arise. Here we focus on how the OC state is destabilized and clarify the dynamical mechanism behind the emergence of the AC state.
Fig.~\ref{fig:delta_p}(a) provides an example of preference evolution with the temperature $\tau=0.26$, slightly larger than the upper threshold $\tau_{c_2}$ of the OC state. The preference is characterized by the probability difference $\Delta p_{{s^{i}}}=p_{{s^{i}},{a_1}}-p_{{s^{i}},{a_2}}$, a positive value means going to the bar is preferred and vice versa. As can be seen, the population initially is of no preference and later self-organizes into the three subgroups --- 50 individuals with  $\Delta p_{{s^{i}}}=1$, 50 with  $\Delta p_{{s^{i}}}=-1$, and 1 with  $\Delta p_{{s^{i}}}\approx 0$ after around $10^3$ steps.
The corresponding PDF of $\Delta p_{{s^{i}}}$ is shown in the left panel of Fig.~\ref{fig:delta_p}(b), showing the strong preference as the two peaks around $\pm 1$ dominate together with a much lower distribution around zero for the pathetic player --- the typical preference profile for the OC state.
However, this state becomes destabilized as time goes by as follows. The preferences of individuals in the two subgroups are not strong enough for the temperature $\tau>\tau_{c_2}$, they deviate by chance from the actions suggested by their Q-tables. Once such action deviation occurs, this could lead to a cascade of action deviation for the rest, including the pathetic player, who may become less random and stick to one of the two choices for some states. As more and more action deviations occur, the average reward $\langle R\rangle$ for some players cannot be guaranteed, which again leads to more deviations, and finally the choice consistency of two subgroups collapses. This is what we see in the latter stage of evolution in Fig.~\ref{fig:delta_p}(a), and the frequency in the right panel of Fig.~\ref{fig:delta_p}(b) shows a blurred distribution between the two peaks $\Delta p_{{s^{i}}}=\pm1$. Though two preferences indicated by the two peaks are still strong.

As $\tau$ is further increased, the OC state is not reachable at all, and the AC state can be seen. This is because the required strong preference profile fails to develop, as observed in Fig.~\ref{fig:delta_p}(c).
The corresponding frequencies of $\Delta p$ shown in Fig.~\ref{fig:delta_p}(d) shows that the weak preference dominates as the profile around zero is much larger. Though, the two peaks around $\Delta p_{{s^{i}}}=\pm1$ are still present and comparable to the middle one.
Although frequencies in Fig.~\ref{fig:delta_p}(d) are statistically symmetrical regarding going or not going to the bar, there is a time-scale separation in weak and strong preference cases that will introduce a bias in superposition.
Specifically, the preference switching for these determined has a much longer time scale than the weak preference cases, see  Appendix~\ref{sec:five}.  Given the slow-varying bias, the superposition of the aggregate of the weak determined actions results in a wider distribution of $A(t)$. This thus leads to enhanced volatility compared to the purely weak determined actions and explains the emergence of anti-coordination.
Obviously, the occurrence of anti-coordination is different from the herding effect as observed in the original MG model~\cite{Challet1997Emergence,Challet2005MinorityGI}. The herding is the consequence of adopting the same lookup tables by many people, while the AC here is due to the strong preference in some states, and time scale separation for the preference switching, the attending number of the slow switching aggregate introduces a bias to the rest of the fast switching aggregate.

Finally, when $\tau\gtrsim 5$, the temperature is so high that no one can hold any strong preference anymore, as the two peaks around $\pm 1$ completely disappear, shown in Fig.~\ref{fig:delta_p}(e) for $\tau= 10$. The corresponding frequency of $\Delta p$ in Fig.~\ref{fig:delta_p}(f) shows a narrow distribution around $\Delta p_{{s^{i}}}=0$. As a result, $\sigma^2/N \rightarrow 0.25$, the evolution approaches the result of the benchmark case as expected.

\section{Phase diagram}\label{sec:phasediagram} 

\begin{figure}
 \centering
\includegraphics[width=0.9\linewidth]{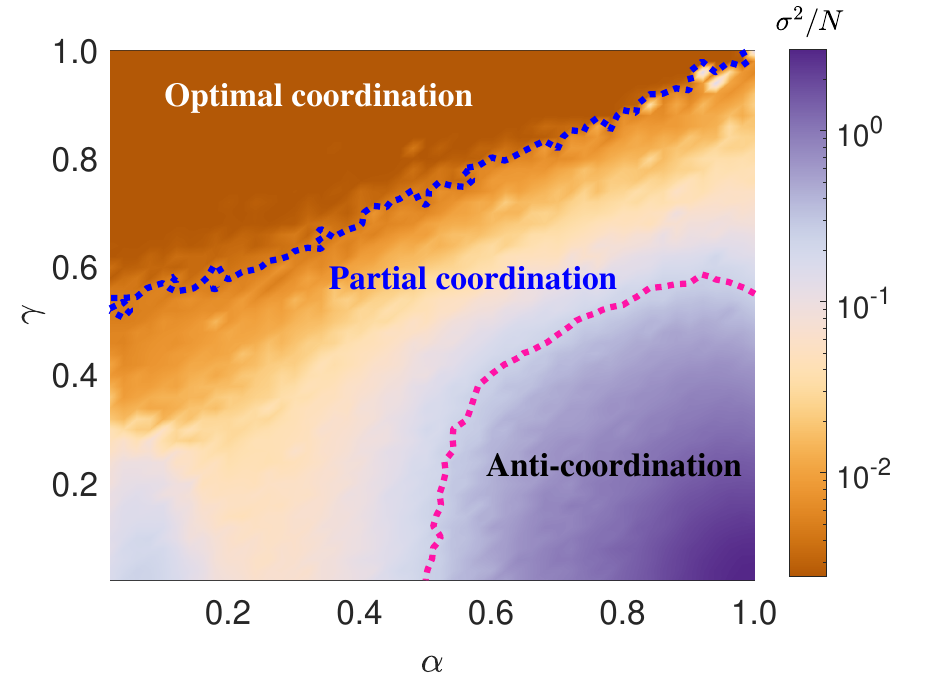}
\caption{\textbf{Phase diagram of the volatility in the learning parameter domain}. 
Three regions are observed: optimal coordination (OC), partial coordination (PC), and anti-coordination (AC). The boundary between OC and PC is by setting the threshold of  $\sigma^2/N = 0.005$, and the other one between PC and AC corresponds to the benchmark value $\sigma^2/N = 0.25$. Each data is averaged over 2000 times after a transient of 8000, and the logarithmic scale is used here for the color coding.
Other parameters: $\tau=0.1$, $N=101$.}
 \label{fig:phasediagram}
 \end{figure}

To systematically examine the impact of the learning parameters, we provide the phase diagram of the volatility in the $\alpha-\gamma$ parameter space by fixing $\tau=0.1$, as shown in Fig.~\ref{fig:phasediagram}. It shows that the domain can be divided into three regions: optimal coordination, partial coordination, and anti-coordination.
As seen, the OC region typically corresponds to the combination of small $\alpha$ and large $\gamma$, which can be interpreted as the individuals care about both the historical experience and the long-term return. The opposite scenario with large $\alpha$ and small $\gamma$ corresponds to the AC where the volatility $\sigma^2/N >0.25$. The PC state is located between these two regions, where the volatility is larger than the value of optimal coordination but smaller than the benchmark value of 0.25.

These observations are in line with our exception. Because once individuals are forgetful (with a large $\alpha$), few lessons can be drawn from history, the poor performance in coordination is expected as the Q-learning loses its strength. Given a small learning rate $\alpha$, the larger the discount factor $\gamma$ is, the stronger guidance there is from the future, which better directs the population into the OC as seen in Fig.~\ref{fig:phasediagram}.
The dependence of performance is qualitatively the same as the previous work in~\cite{Zheng2024decoding, Ding2023emergence}, where a high prevalence of trust or cooperation is observed when the Q-learning individuals are of small $\alpha$ and large $\gamma$.

 \begin{figure}[t]
\centering
\includegraphics[width=0.8\linewidth]{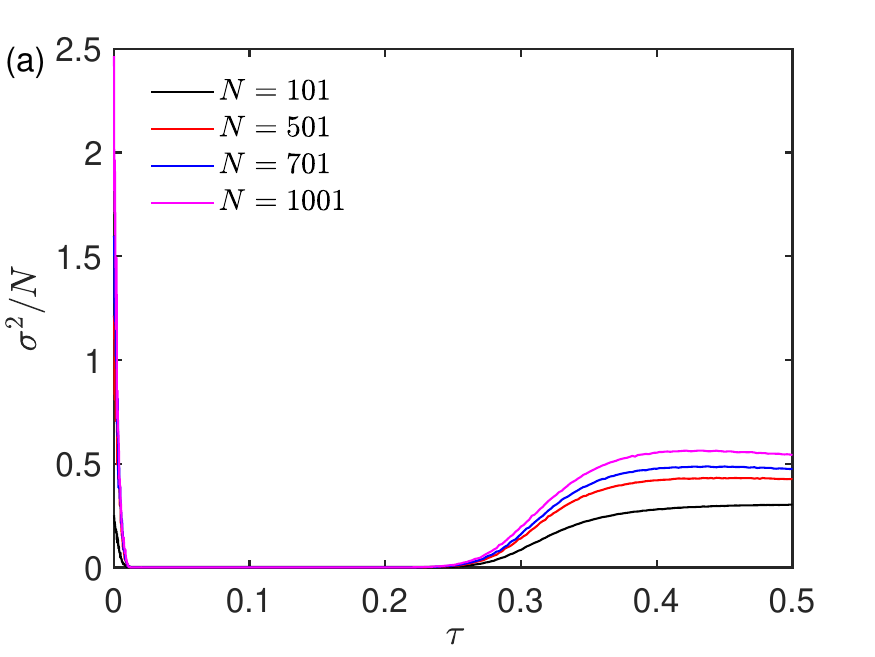}\\
\includegraphics[width=0.8\linewidth]{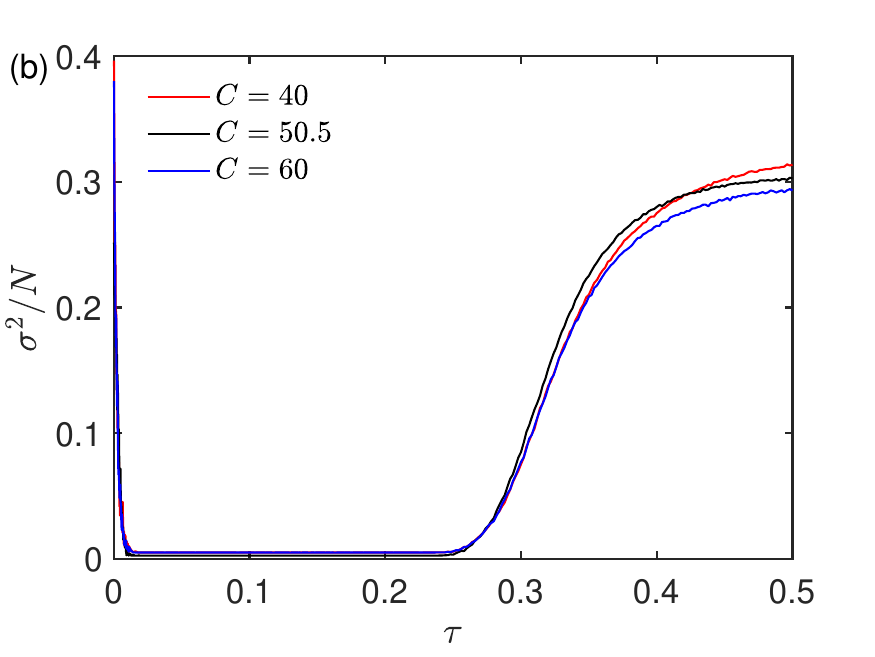}
\caption{\textbf{Volatility as a function of the temperature $\tau$ for different population sizes (a) and bar capacities (b)}. 
Each data is averaged over 100 realizations, and for $5\times10^3$ time average after a transient of $5\times10^4$. $C=N/2$ in (a), and $N=101$ in (b).
Other parameters: $\alpha=0.1$,$\gamma=0.9$.
}
 \label{fig:robustness}
 \end{figure}

 \section{Robustness}\label{sec:Robustnes} 

 The observations made in population size $N=101$ and the bar capacity $C=N/2$ are robust when the different population sizes and bar capacities are varied. Fig.~\ref{fig:robustness}(a) shows the results when the population size is increased up to $1001$, the phase transition and the regions for different phases remain almost unchanged. Similar robustness is seen in Fig.~\ref{fig:robustness}(b) when the bar capacity $C$ is varied by fixing $N=101$. We see that the capacity $C$ of the bar is not necessarily half of the population, a setup that is often adopted in the study of Minority Game.

Furthermore, these observations can also be seen in an even number of population $N=2m$, where $m$ is an integer. In this case, one needs to make a rule that determines which side wins, i.e. the winning criterion regarding the capacity $C$. We find that when the bar capacity $C=m$, the observations remain the same as the above odd number scenario. In particular, the orchestration of OC state is slightly different, where the two frozen subgroups are not strictly of the same size anymore. One pathetic player emerges that will always be the loser, $m$ of the rest prefer one action, $m-1$ players choose the opposite. An example of $N=100$ is shown in  Appendix~\ref{sec:six}.

\section{Discussion}\label{sec:Discussion} 

Given the fundamental importance of efficient resource allocation and the limitations of static solutions in previous studies,
here in this work, we provide a new solution to the problem by shifting the paradigm to reinforcement learning.
Specifically, we adopt a Q-learning algorithm to solve the Minority Game, where each player is empowered with an evolving Q-table that guides one's move. We reveal that the population could evolve into different phases, depending on the trade-off between their exploitation of Q-tables and the exploration.  On the one hand, the population with insufficient exploration is prone to fall into periodic states, many of which are metastable, and the coordination is suboptimal. With too much exploration, on the other hand, the coordination is also bad since the individuals act just like flipping the coin. To our surprise,  when the trade-off between the two is balanced, we observe the emergence of optimal coordination, where the volatility is minimized around the capacity. 
Mechanistically, there is a symmetry-breaking within the population's preference, where nearly half of the people invariably go to the bar, the other half never do for the given state, and there is one ever-losing/irresolute individual who restlessly switches its action. Between the case of optimal coordination and exploration-only, there is an anti-coordination phase where the coordination is even worse than the benchmark scenario where the decision is made by simply flipping the coin. This is due to the bias introduced by those individuals with a strong preference in some states.

Interestingly, in an early work~\cite{Whitehead2008farol}, where the EI Farol bar problem is revisited by an RL algorithm~\cite{Erev1998predicting}, this work predicted with a mathematical proof that the population is going to be subdivided into two groups: those who invariably go to the bar and those who never do. Our work thus validates the sorting phenomenon, though the model setups are different. In another relevant work~\cite{ren2007randomness}, they also identified the optimal amount of randomness in the promotion of cooperation prevalence with imitation learning; this resonance-like phenomenon is similar to the OC region in Fig.~\ref{fig:tau}(a), though our work studies a completely different question, and the paradigms adopted are also different.  
We have to stress that our RL approach is fundamentally distinct from most prior works with handcrafted rules. A recipe by design in those works, e.g. in Ref.~\cite{Challet1997Emergence} is composed of a fixed number of strategies, but it's hard to expect in real life that players have the same number of rules sampled from a common pool.  Moreover, the strategies themselves within the recipe are assumed to be unchanged, which also fail to capture the adaptive nature of strategy in the real world.

As scarcity is an intrinsic property of our world, the proposed paradigm provides a novel perspective to solve the resource allocation problems in our society.  Our findings suggest that by integrating the experience, the present reward, and the expected reward in the future, the population can reach optimal coordination during the process where the individuals seek to maximize their accumulated rewards. This reconciles the individuals' self-interests and the collective resource utilization.
Our results thus may also provide a plausible solution to the efficient market assumption~\cite{Samuelson2005Economics}, and its failure may be attributed to the balance-breaking of the trade-off between exploitation and exploration.
Compared to the original scheme in Ref.~\cite{Challet1997Emergence,Challet2005MinorityGI}, our work shows that the Q-table as the policy is learnable, and is coevolving with the environment. In particular, to achieve optimal coordination, their Q-tables self-organize into structured sorting. It is the policy heterogeneity that makes optimal coordination possible. This heterogeneity spontaneously emerges in our Q-learning scheme but needs to be artificially tuned in Ref.~\cite{Challet1997Emergence,Challet2005MinorityGI}.

It's important to delimit the limitations of our findings.
The major limitation is that our results were obtained for the Minority Game, which is a toy model for resource allocation issues. However, this model has been readily modified and generalized to incorporate many new ingredients~\cite{chakraborti2015statistical}. It would be possible and interesting to expand our findings to some realistic scenarios, such as Kolkata Paise Restaurant problem~\cite{chakrabarti2009kolkata}, the stable marriage problem~\cite{fenoaltea2021stable}, etc. A more substantial step comes from behavioral experiments~\cite{camerer2011behavioral}, it would be desired to see such experiments to be carried out to validate or falsify our findings and to further compare the realistic processes to the rich spectrum of dynamics uncovered here. In addition, it's also interesting to see the possible application of our RL framework to understand the resource allocation problems in other contexts, such as the allocation of nutrients~\cite{erickson2017global} and enzymes~\cite{giunta2022optimal} in biology.
The second limitation is that we have not yet developed any theoretical treatment for the dynamical process in our reinforcement learning paradigm.  A technical difficulty comes from mapping the discrete Q-table into some continuous variables to further borrow the approaches in statistical mechanics~\cite{Challet2005MinorityGI}. Even so, the stability dependence of periodic states on the temperature-like parameter suggests the energy landscape theory~\cite{Wales2018exploring, Fang2019nonequilibrium} may be a good starting point, and we leave it to the future.

\section*{Data availability}
All data are available at \href{https://github.com/chenli-lab/RL-Minority-game/tree/main/data}{https://github.com/chenli-lab/RL-Minority-game/tree/main/data}.

\section*{Code availability}
The code for generating key results in this study is available at \href{https://github.com/chenli-lab/RL-Minority-game/tree/main/code}{https://github.com/chenli-lab/RL-Minority-game/tree/main/code}.

\section*{ACKNOWLEDGEMENTS}
This work is supported by the National Natural Science Foundation of China (Grants Nos. 12075144, 12165014), the Fundamental Research Funds for the Central Universities (Grant No. GK202401002), and the Key Research and Development Program of Ningxia Province in China (Grant No. 2021BEB04032).

\appendix
\section{The evolution of procedure}\label{sec:one}

\begin{figure}[htbp]
 \centering
\includegraphics[width=1.0\linewidth]{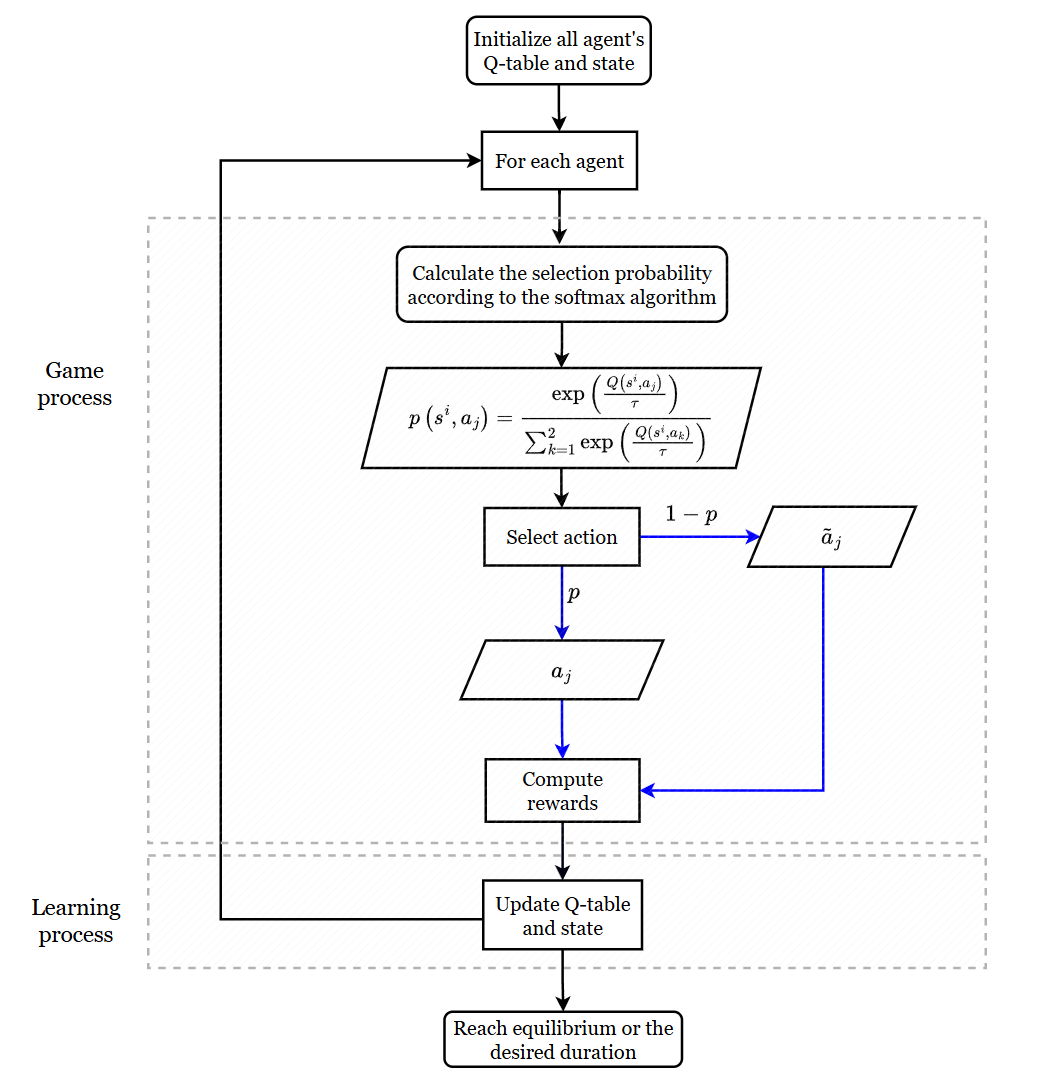}
\caption{ Protocol flowchart for the Minority Game, consisting of two main parts: game process and learning process. The probability calculation in the flowchart is according to Eq.~(\ref{eq:probability}) in the main text. Subsequently, individuals choose action $a_j$ with probability $p$ and choose the opposite action $\tilde{a}_j$ otherwise.}
 \label{fig:protocol}
 \end{figure}

The setup of the Minority Game, outlined in Fig.~\ref{fig:protocol}, employs a synchronous updating: (1) Initialize Q-tables and actions: set all Q-tables to 0, representing agents’ initial unawareness of the environment. Assign a random initial action to each agent: either go to the bar ($a_1$) or not go to the bar ($a_2$). (2) Game process: each agent selects an action based on the softmax-calculated probability. Then, they calculate their rewards based on the outcome. (3) Learning process: agents update their Q-tables and states. Repeat steps 2 and 3 until the system reaches statistical stability or the predefined time limit.

\section{An evolving pattern towards optimal coordination}\label{sec:two}

\begin{figure*}[btph]
 \centering
\includegraphics[width=0.75\linewidth]{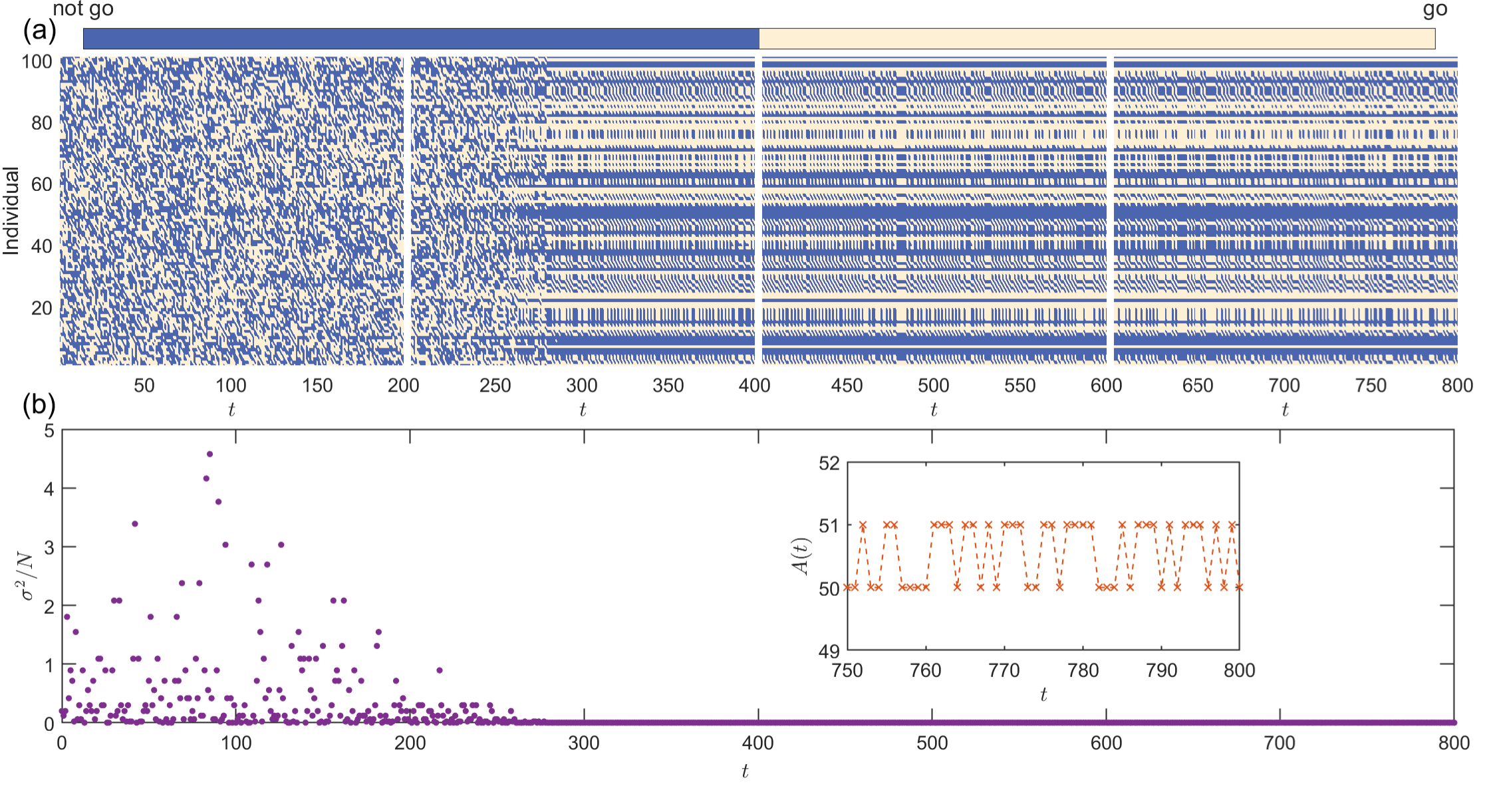}
\caption{The evolution towards optimal coordination. (a) The spatial-temporal patterns of action with the temperature $\tau=0.1$, consisting of four typical time windows (from left to right): 0$\sim$200, 200$\sim$400, 400$\sim$600, 600$\sim$800. (b) The corresponding time series of the system volatility, where the inset is a time series for the number of people who went to the bar $A(t)$ by enlarging the time window during 750$\sim$800. Parameters: $\alpha=0.1$, $\gamma=0.9$, $N=101$.}
 \label{fig:optimal}
 \end{figure*}

From the pattern in Fig.~\ref{fig:optimal}(a) above, the evolution process is more intuitively understood. As can be seen from the pattern, in the initial stage around 0$\sim$260 steps, actions tend to be selected randomly. As the gap between Q-values widens and gradually becomes stable, the actions of most individuals become distinct, that is, they have been determined to choose to go or not to go in different states. Fig.~\ref{fig:optimal}(b) reports the corresponding time series of the volatility. As seen, after 260 steps, the value is minimized, meaning the system is optimally coordinated. The insert shows the evolution of the number of people going to the bar for the typical 50 steps within the steady state, $A(t)$ randomly jumps between 50 and 51, which is different from the periodic jumping shown in Fig.~\ref{fig:exploitation_only}(b) in the main text.

\section{The preference fact within the OC state}\label{sec:three}

\begin{figure}[htbp]
 \centering
\includegraphics[width=0.85\linewidth]{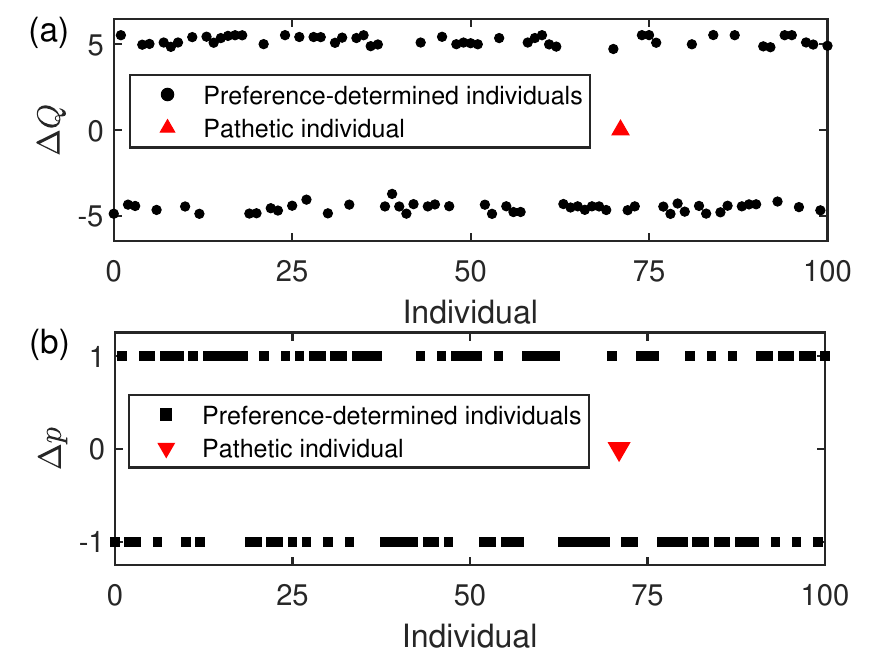}
c\caption{Q-value differences and the probability differences within $s(t)=50$ corresponding to the last step in Fig.~\ref{fig:OC}(a) shown in the main text. (a) The $\Delta Q=Q_{50,a_1}-Q_{50,a_2}$ at the final time for all $N=101$ individuals. Player 71 as the pathetic individual is represented by a red triangle, while the remaining $N-1$ individuals are preference-determined and represented by black circles. (b) The $\Delta p=p_{50,a_1}-p_{50,a_2}$ for all $N=101$ individuals. Individual-71 is represented by a red triangle, while the remaining $N-1$ individuals are represented by squares. The setup is the same as Fig.~\ref{fig:OC} in the main text.}
 \label{fig:note3}
 \end{figure}

To illustrate the preference property within the optimal coordination state more clearly, we provide the Q-value difference and the difference of the two preference probabilities within state $s=50$, corresponding to the last step in Fig.~\ref{fig:OC}(a) in the main text.
In Fig.~\ref{fig:note3}(a), $\Delta Q=Q_{s_{50},a_1}- Q_{s_{50},a_2}$ indicates that if $\Delta Q>0$, individuals are inclined to go to the bar; otherwise, they prefer not to. It can be observed that, except for individual 71, whose $\Delta Q \approx 0$, the preferences of all other individuals are determined. In Fig.~\ref{fig:note3}(b), the corresponding $\Delta p=p_{s_{50},a_1}- p_{s_{50},a_2}$ further gives the difference between the probability of choosing to go to the bar and the probability of not going. Similarly, aside from individual 71, who has $\Delta p \approx 0$ (thus randomly wandering between going and not going), the probability differences for all other individuals are either 1 (certainly to go to the bar) or -1 (certainly not to go). 

\section{The competition process before reaching OC state}\label{sec:four}

\begin{figure*}[htbp]
 \centering
\includegraphics[width=0.8\linewidth]{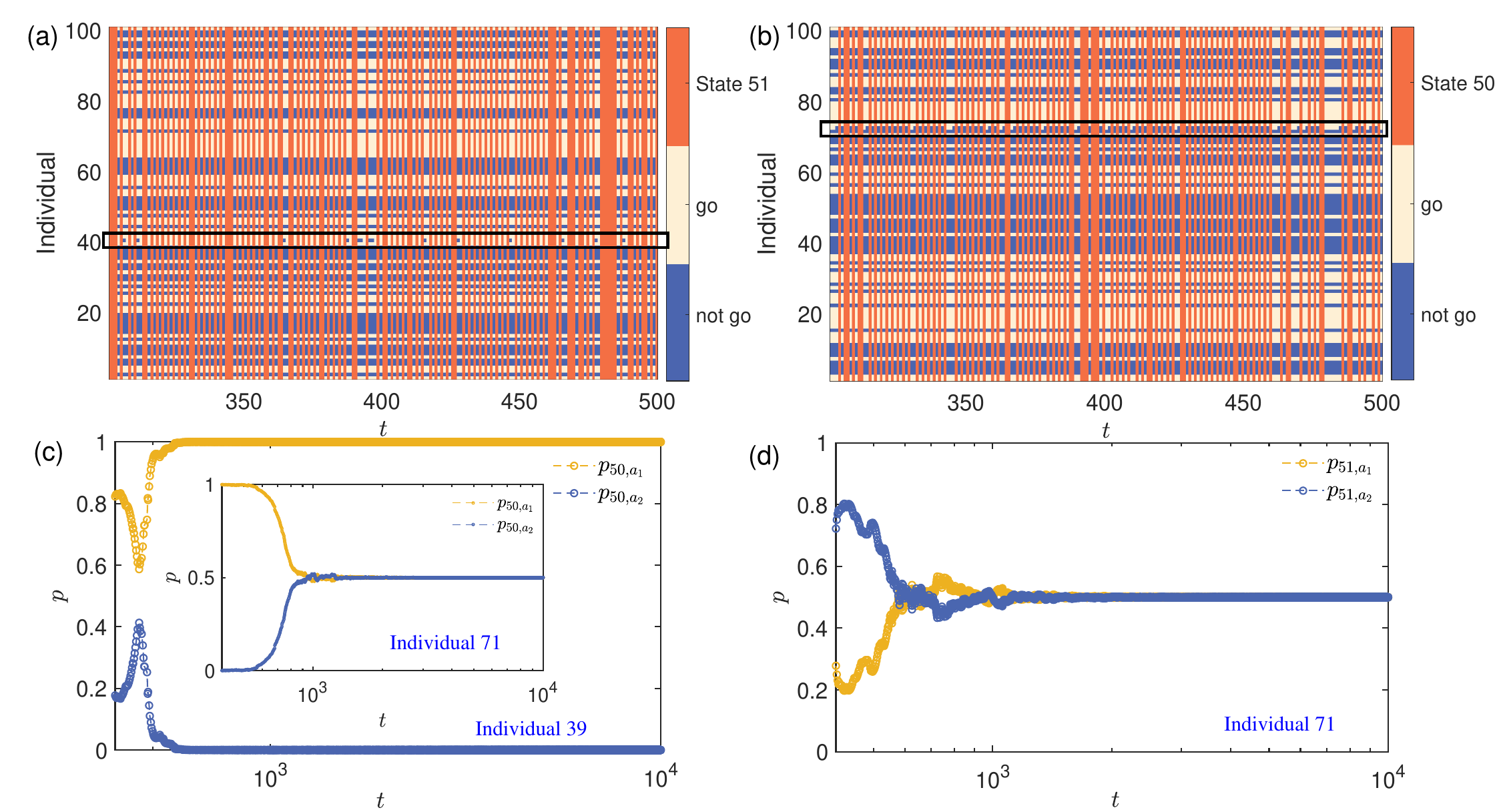}
\caption{The evolution process before the arrival of the OC state. The patterns in $s(t)=50$ (a) and $s(t)=51$ (b). Besides the two actions ($a_1=$``go” and $a_2=$``not go”), being within the other state is color-coded with orange. In $s(t)=50$, individual 39 indicated by a rectangle emerges as the irresolute one who randomly switches its action, whereas all other individuals’ actions remain unchanged all the time. While individual-71 indicated by a rectangle becomes the irresolute one who randomly switches its action in $s(t)=51$. (c) The temporal evolution of individual-39's preference $p$ for the two actions within state 50, and individual 39 finally becomes determined to go to the bar (i.e. $p_{50,a_1}\rightarrow 1$). The inset shows that after individual 39 becomes the determined player, individual 71 is forced to be the pathetic individual in state 50. (d) The temporal evolution of individual 71's preference $p$ for the two actions in state 51. 
Other parameters: $\tau=0.1$, $N=101$.
}
 \label{fig:Comparable}
 \end{figure*}

Fig.~\ref{fig:Comparable} reports the nontrivial dynamical process before the OC state is reached.  
The patterns in Fig.~\ref{fig:Comparable}(a) and (b) correspond to the cases in $s(t)=$50 and 51, respectively. In both of these states, we observe the occurrence of symmetry breaking leading to the formation of two pathetic individuals -- an individual labeled ``39" in state 50 and another individual labeled ``71" in state 51. This symmetry breaking in states 50 or 51 results in a fixed pattern wherein 50 individuals opt to go to the bar, while the remaining 50 individuals persistently choose not to. 

The competition comes from the following situations: 1) within $s(t)=50$, individual 39, always obtains a zero reward as the pathetic player regardless of his/her choice. This is the same for individual 71 within $s(t)=51$. 2) However, the move of individual 39 (71) (going or not going to the bar) matters for the reward of individual 71 (39) within $s(t)=50 (51)$ as he/she is a determined player within this state. For example, within $s(t)=50$ the choice of not going to the bar for individual 39 makes individual 71 win as he/she is determined to go to the bar in this simulation. Within $s(t)=51$ the choice of going to the bar for individual 71 makes individual 39 win instead as he/she is determined not to go in this state.
Due to the stochastic nature of the simulation, the winning times of the two players (39 and 71) are not equal. In this simulation, individual 39 more frequently finds itself in the minority compared to individual 71, leading to a faster increase in Q-values. This further leads to the formation of preference for individual 39 within $s(t)=51$, and instead, individual 71 becomes the pathetic loser as he/she is the least rewarded player, as shown in Fig.~\ref{fig:Comparable}(c).
In brief, the competition occurs within state $s(t)=50$, the relative advantage in the winning number makes individual 39 escape from being the pathetic loser, and the least rewarded player (individual 71) becomes the ever-losing player within both states finally. 

\section{Time scale separation in the AC state}\label{sec:five}

 \begin{figure*}[htbp]
\centering
\includegraphics[width=0.8\linewidth]{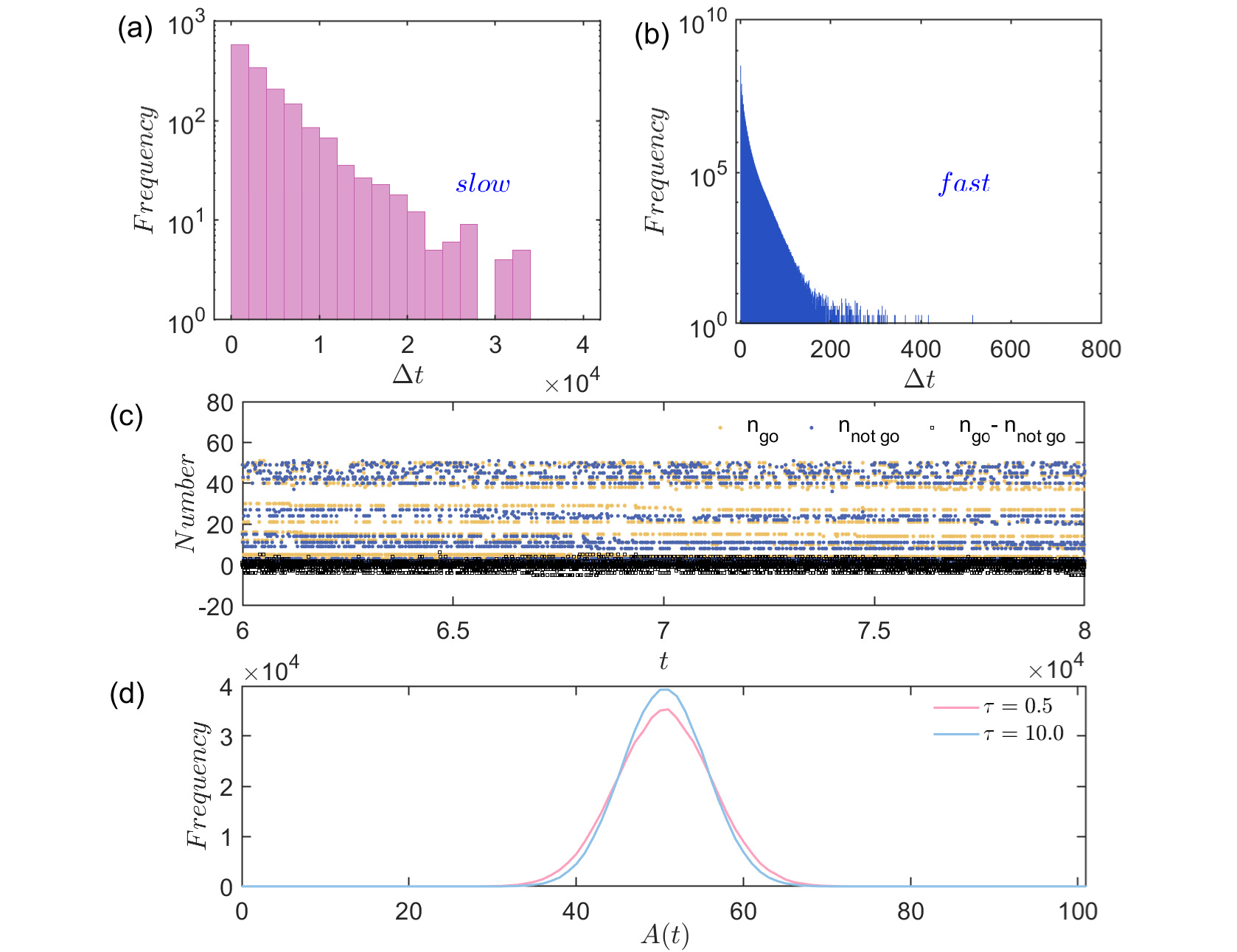}
\caption{Time scale separation in the AC state. Frequency distribution of $\Delta t$ for slow (a) and fast variables (b) within $\tau=0.5$. 
(c) The number of players holding the two strong preferences, going or not going to the bar, and the number difference.   
(d) Frequency distribution for the number of people going to the bar $A(t)$ for $\tau=0.5$ and 10.
Other parameters: $\alpha=0.1$, $\gamma=0.9$, $N=101$.
}
 \label{fig:TimeScale}
 \end{figure*}

According to the temporal evolution of preference $\Delta p_{s^i}$ depicted in Fig.~\ref{fig:delta_p}(c) in the main text, we categorize $-0.3 \textless \Delta p_{s^i} \textless 0.3 $ as the weak preference — individuals exhibiting weak preferences in two actions (``go” and ``not go”). $\Delta p_{s^i} \textgreater 0.9$ or $\Delta p_{s^i} \textless -0.9$ is classified as the strong preference — individuals demonstrating relatively deterministic preferences in their action choices. 
Then, we define $\Delta t$ as the time it takes from strong to weak preference, which corresponds to the action change for those with strong preference, with results being shown in Fig.~\ref{fig:TimeScale}(a). For those with weak preferences, we compute the transition time for the action changes (i.e. from ``go" to ``to go" or the opposite) as $\Delta t$, shown in Fig.~\ref{fig:TimeScale}(b). By comparison, there is a time-scale separation for the two transition times, the transition time for those with a strong preference is two orders of magnitude larger, meaning that once the players hold a strong preference, they stick to them for quite a long time.  

Since the numbers of players holding two strong preferences ($\Delta p=1$ and -1) are not the same, some background always remains, as shown in Fig.~\ref{fig:TimeScale}(c). 
As a result, the remaining background acts as a quasi-static ``substrate", working as a bias to widen the distribution of the attending number of people shown in Fig.~\ref{fig:TimeScale}(d).

 \section{The optimal coordination in an even number of population}\label{sec:six}
\begin{figure*}[htbp]
\centering
\includegraphics[width=0.8\linewidth]{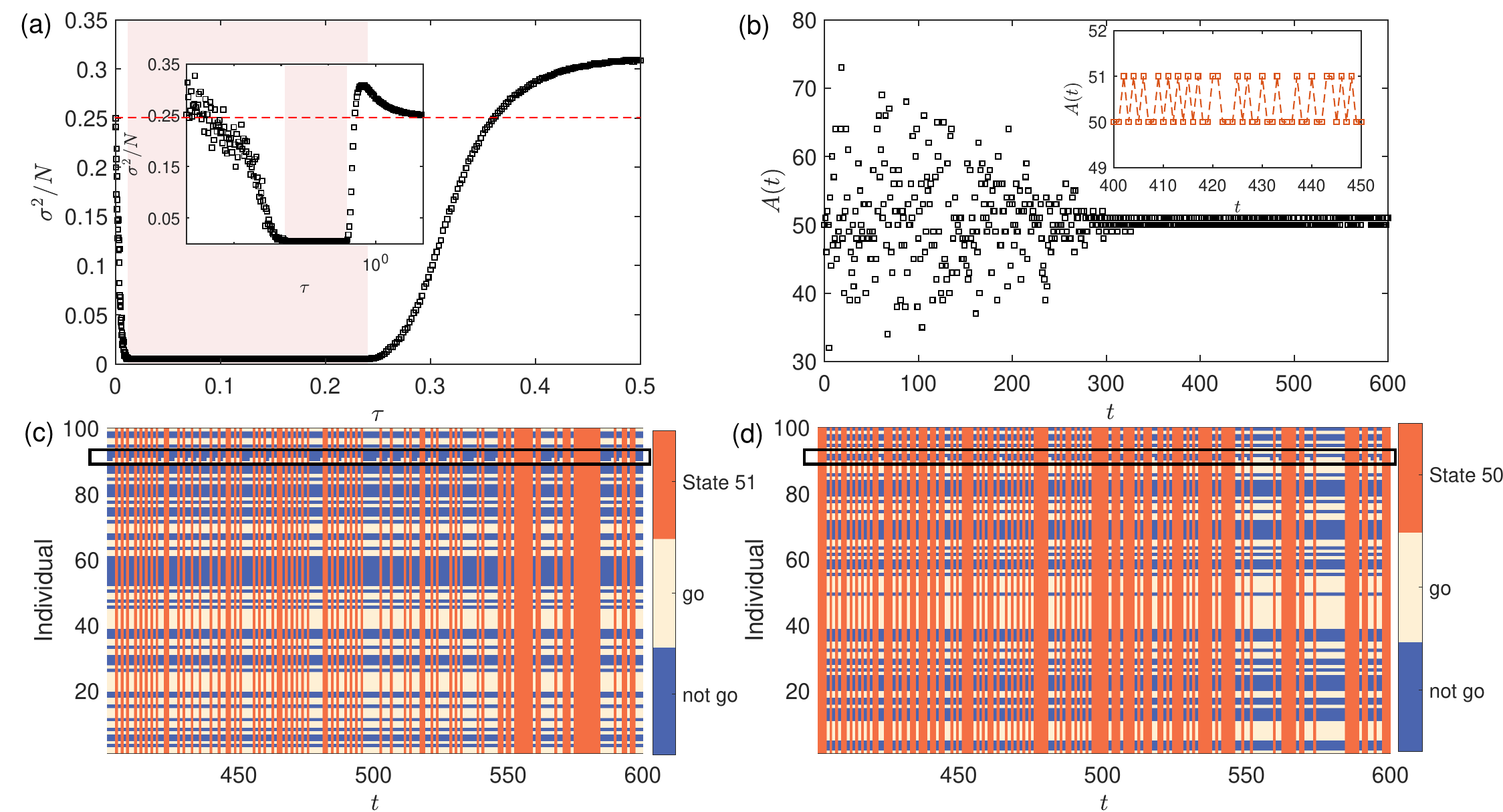}
\caption{The evolution of an even number of population with $N=100$. 
(a) The volatility as a function of the temperature $\tau$. Each data is averaged 100 realizations and for $5\times10^3$
time average after a transient of $5\times10^4$. The inset shows the plot with a logarithmic operation of the x-axis, the shadowed zone corresponds to the region where the optimal coordination is achieved. The red dashed line ($\sigma^2/N$ = 0.25) corresponds to a benchmark scenario where individuals decide whether to go to the bar by simply flipping the coin. 
(b) The typical time series for the number of people going to the bar $A(t)$ by fixing $\tau=0.1$, and the inset is shown by enlarging the time window 400$\sim$450. 
(c,d) The stationary patterns within the optimal coordination state, respectively in $s(t)=50$ (c) and $s(t)=51$ (d).
Other parameters: $\tau=0.1$, $\alpha=0.1$, $\gamma=0.9$.
}
 \label{fig:EvenNumber}
 \end{figure*}

 Fig.~\ref{fig:EvenNumber} shows the case in an even population of $N=100$. The winning criterion is for those whose part ends up with the number $n_{winning}\le C=50$. The volatility as a function of the temperature $\tau$ is shown in Fig.~\ref{fig:EvenNumber}(a), which is almost the same as the result of odd size $N=101$ [Fig. 1(a) shown in the main text]. By fixing $\tau=0.1$, the time series of the attending number $A(t)$ is shown in Fig.~\ref{fig:EvenNumber}(b), where the value of $A(t)$ fluctuates around $C$ in the 50--51 manner. The inset in Fig.~\ref{fig:EvenNumber}(b) clearly shows $A(t)$ random jump between 50 and 51. The patterns in Fig.~\ref{fig:EvenNumber}(c) and (d) correspond to the cases in states $s(t)=$ 50 and 51, respectively. We can observe that one pathetic player emerges as the ever-losing loser, and others have decided to go to or not go to the bar in both states. 
Note that if the winning criterion is changed into $n_{winning}<50$, the observations remain almost the same, except the optimal state now becomes 49-50.

\bibliography{MG}

\end{document}